\documentclass[aps,prc,twocolumn]{revtex4}
\usepackage{pdfpages}
\usepackage{graphicx}
\usepackage{amsmath}
\usepackage{booktabs}
\usepackage{multirow}
\usepackage{setspace}
\usepackage{color}
\begin{document}

\title{Heavy dibaryons $\Xi^{(*)}_{cc}\Xi^{(*)}_{cc}$ and $\Xi^{(*)}_{bb}\Xi^{(*)}_{bb}$}

\author{An-Su Lu, Mao-Jun Yan${\footnote{yanmj0789@swu.edu.cn}}$, Chun-Sheng An${\footnote{ancs@swu.edu.cn}}$, and Cheng-Rong Deng${\footnote{crdeng@swu.edu.cn}}$}
\affiliation{$^a$School of Physical Science and Technology,
Southwest University, Chongqing 400715, China}

\begin{abstract}
We systematically investigate the dibaryons $\Xi^{(*)}_{cc}\Xi^{(*)}_{cc}$ (di-$\Xi_{cc}$) and
$\Xi^{(*)}_{bb}\Xi^{(*)}_{bb}$ (di-$\Xi_{bb}$), with various isospin-spin configurations $I(J^P)$ in a
nonrelativistic quark model. For the di-$\Xi_{cc}$ system, only the single channels $\Xi_{cc}\Xi^*_{cc}$ and
$\Xi^*_{cc}\Xi^*_{cc}$ with $0(1^+)$ are capable of forming deuteronlike bound states, with the $\sigma$
meson exchange playing a decisive role. Those states have binding energies of approximately $-1.5$ MeV and $-3.3$
MeV and sizes of 2.37 fm and 1.87 fm, respectively. The coupled channel effect in the di-$\Xi_{cc}$ system
with $0(1^+)$ enhances the attraction. As a result, this di-$\Xi_{cc}$ system can establish a deuteronlike
configuration, with the binding energy of $-7.5$ MeV relative to the threshold $\Xi_{cc}\Xi_{cc}$ and the
size of approximately 1.40 fm. For the di-$\Xi_{bb}$ system, the single channels with $0(1^+)$, $0(2^+)$, and
$0(3^+)$ can give rise to deuteronlike bound states with binding energies ranging from $-6.1$ MeV to $-14.3$ MeV.
Additionally, the di-$\Xi_{bb}$ system with $1(0^+)$ and $1(2^+)$ can also establish deuteronlike bound states
with binding energies of around $-0.5$ MeV. When considering the coupled channel effect in the di-$\Xi_{bb}$
system with $0(1^+)$, a compact hexaquark state is formed, exhibiting a binding energy of $-21.2$ MeV relative
to the threshold $\Xi_{bb}\Xi_{bb}$ and a size of 0.53 fm. In this state, the $\pi$ meson exchange provides a
very powerful attractive force. The meson exchange interactions in the quark model is dispensable in the
di-$\Xi_{bb}$ bound states, except for $\Xi_{bb}^*\Xi_{bb}^*$ with $1(0^+)$.

\end{abstract}
\maketitle

\section{introduction}

A dibaryon is defined as a system with a baryon number $B=2$, composed of six valence quarks.
These quarks can manifest in two forms: either as a loosely baryon-baryon configuration or
as a tightly bound, spatially compact hexaquark state. The best-known dibaryon, the deuteron,
is a loosely bound system composed of a proton and a neutron~\cite{DeVries:1987atn}, whereas
the state $d^*(2380)$ is widely regarded as a compact hexaquark state because of its large binding
energy relative to the threshold $\Delta\Delta$~\cite{Clement:2016vnl,Clement:2020mab,Dong:2023xdi}.
Alternatively, the state $d^*(2380)$ is described as a $\mathcal{D}_{12}\pi$ configuration with
the pion in relative P-wave~\cite{Gal:2013dca,Gal:2014zia}.
Deuteron-like hadronic molecules have emerged as the leading interpretation for the
internal structure of the so-called $XYZ$, $P_c$, $P_{cs}$, $T_{cc}$ and $T_{cs}$ states~\cite{Dong:2017gaw,Guo:2017jvc,Dong:2021bvy,Dong:2021juy,Chen:2022asf,Meng:2022ozq,Huang:2023jec,Liu:2024uxn}.
It is therefore natural to explore similar
dibaryons~\cite{Carames:2015sya,Meng:2017fwb,Yang:2019rgw,Chen:2018pzd,Junnarkar:2019equ,
Wang:2019gal,Lyu:2021qsh,Deng:2023zlx,Wang:2024riu,Junnarkar:2024kwd,Zhang:2025zaa,Cui:2025qpm,
Xing:2025uai,Mathur:2022ovu}, which contain one or more heavy quarks in their valence structure, and to
search for their discovery when the requisite center-of-momentum energy and luminosity
become accessible at experimental facilities.

The Heavy Diquark-Antiquark Symmetry (HADS) posits that the heavy quark pair $[QQ]$,
confined in the color $\bar{\mathbf{3}}$ state within doubly heavy baryons, forms
a compact object due to the strong attractive color Coulomb interaction~\cite{Savage:1990di}.
In the heavy quark limit, the heavy diquark $[QQ]$ behaves as a point-like particle, acting
purely as a static color source in the $\bar{\mathbf{3}}$ channel. The HADS facilitates
the establishment of a correlation between the doubly heavy baryons $QQq$ ($\bar{Q}\bar{Q}\bar{q}$)
and singly heavy mesons $\bar{Q}q$ ($Q\bar{q}$). Building on the molecular assumption
for the $P_c$ states, the existences of bound tri-charmed dibaryons, $\Xi_{cc}^{(*)}\Sigma_c^{(*)}$,
as well as the hexaquark molecule composed of a tetraquark state $T_{\bar{c}\bar{c}}$
and $\bar{D}^*$ have been predicted within the HADS framework~\cite{Liu:2018zzu,Pan:2019skd,Pan:2020xek,Pan:2022whr}.

The characteristic size of the doubly charmed state $T^+_{cc}(3875)$ reported by the LHCb
Collaboration suggests its deuteron-like molecular configuration, specifically
$DD^*$~\cite{LHCb:2021vvq,LHCb:2021auc}. In the wake of this discovery, theorists proposed a
wide range of possible tetraquark partners within various theoretical frameworks~\cite{Mathur:2021gqn,
Chen:2021vhg,Deng:2021gnb,Karliner:2021wju,Albaladejo:2021vln,Deng:2022cld,Li:2023hpk,Aoki:2023nzp,
Alexandrou:2023cqg,Padmanath:2023rdu,Liu:2023hrz,Lin:2023gzm,Radhakrishnan:2024ihu,Chen:2024snh,
Tanaka:2024siw,He:2025hhh}. According to the HADS, the dibaryon partners,
${\bar\Xi}^{(*)}_{cc}\bar{\Xi}^{(*)}_{cc}$ (which is equivalent to $\Xi^{(*)}_{cc}\Xi^{(*)}_{cc}$), which are
derived by replacing a $c$-quark in the state $T^+_{cc}(3875)$ with a diquark $[\bar{c}\bar{c}]$,
could produce a bound deuteron-like dibaryon.

According to the heavy quark flavor symmetry (HQFS), the observation of $T^+_{cc}(3875)$ suggests that
its bottom counterpart, $T^-_{bb}$ with $01^+$, should exist. The existing studies indicated that
the state $T^-_{bb}$ with $01^+$ is more stable bound state compared to $T^+_{cc}(3875)$~\cite{Deng:2021gnb,Deng:2018kly}.
Furthermore, the double bottom baryon $\Xi_{bb}$ is anticipated to be observed in the near future~\cite{Ali:2018xfq}.
However, the production of such states in experiments poses significant challenges.
Similar to the state di-$\Xi_{cc}$, the di-$\Xi_{bb}$ configuration may also give rise
to the formation of bound states. Although the doubly bottom baryon $\Xi^{(*)}_{bb}$ has
yet to be observed experimentally, its theoretical predictions suggest intriguing possibilities
for future discoveries in the field of hadronic physics.

Experimental data concerning interactions between two heavy baryons
is quite limited due to the challenges associated with generating two heavy baryons
in laboratory settings. On the lattice quantum chromodynamics (LQCD) side, several
heavy flavor dibaryon states have been studied~\cite{Junnarkar:2019equ,Lyu:2021qsh,
Mathur:2022ovu,Junnarkar:2024kwd,Xing:2025uai}, including di-$\Omega_{ccc}$, di-$\Omega_{bbb}$,
di-$\Lambda_{c}$, di-$\Sigma_{c}$, $\Sigma_c\Xi_{cc}$, $\Omega_{cc}\Omega_{c}$, $\Omega_{bb}\Omega_{b}$,
and $\Omega_{cbb}\Omega_{ccb}$. These studies have predicted the existence of several
bound states within these systems. However, it is noteworthy that the dibaryon states
di-$\Xi_{cc}$ and di-$\Xi_{bb}$ remain under explored in LQCD research. Up to now, only
the one-boson-exchange (OBE) model has explored the state di-$\Xi_{cc}$, predicting the
existence of several potential bound states~\cite{Meng:2017fwb,Yang:2019rgw}. In contrast,
there has been no any theoretical researches on the state di-$\Xi_{bb}$.

In this study, we aim to conduct a systematic investigation of the properties of the heavy dibaryons
di-$\Xi_{cc}$ and di-$\Xi_{bb}$. We will analyze a range of isospin-spin configurations in order to
identify potential bound states within the framework of the quark model. We will investigate
their binding energies, underlying binding mechanisms, and spatial structures. With this model,
our previous work has investigated the natures of the single flavored dibaryon di-$\Delta$, di-$\Omega$,
di-$\Omega_{ccc}$, and di-$\Omega_{bbb}$~\cite{Deng:2023zlx}. The primary objective of this work
is to deepen our understanding of the properties and structures of heavy dibaryon states from
a phenomenological perspective. We aim to provide valuable insights that can inform future
experimental endeavors aimed at identifying these dibaryon states. In addition, we will investigate
the fundamental binding mechanisms that contribute to the formation and stability of these bound
states. Our research seeks to uncover evidence for binding mechanisms that are unique to these
systems, thereby enhancing our comprehension of the dynamics of heavy quarks and the meson-exchange
interactions among light quarks. This exploration will yield important insights into the fundamental
forces that govern the behavior of heavy dibaryons.

Following the introduction, the paper is organized as follows: In Sec. II, we present
the quark model Hamiltonian. In Sec. III, we provide a brief overview of the wave functions
for ground-state dibaryons. In Sec. IV, we present the numerical results and discussions.
Finally, the paper concludes with a brief summary in Sec. V.

\section{Theoretical frameworks}

Quantum Chromodynamics (QCD) is widely regarded as the fundamental theory of strong
interaction, governing the dynamics of quarks and gluons. At the hadron scale, QCD exhibits
highly nonperturbative behavior due to the intricate infrared dynamics of the non-Abelian
SU(3) gauge group. Direct calculations of hadron spectra and hadron-hadron interactions
from first principles in QCD remain extremely challenging. As a result, a less rigorous
yet highly effective approach the QCD-inspired quark models serve as a powerful
tool for gaining physical insight into these complex strong-interaction systems.
The quark models are based on the assumption that hadrons are color-singlet,
nonrelativistic bound states of constituent quarks, with phenomenologically
derived effective masses and interactions. These models were developed based
on a comprehensive understanding of the nature of baryons and the nucleon-nucleon
interactions.

The naive quark model (Isgur-Karl model) typically includes an effective one-gluon-exchange (OGE)
potential, $V^{\rm oge}$, which directly stems from the one-gluon exchange
diagram in QCD~\cite{DeRujula:1975qlm}, and an artificial quark confinement
potential, $V^{\rm con}$. This model offers an excellent description of light
baryons~\cite{Isgur:1978xj,Isgur:1979be}. In the context of nucleon-nucleon
interactions, the model can capture the short-range repulsive core via the spin-spin
component of the inter-quark interaction between nucleons, as well as the Pauli
exclusion principle, which is enforced by the quark structure of the nucleon~\cite{Liberman:1977qs}.
However, the model does not account for the medium-range attraction~\cite{Neudachin:1977vt,Fujiwara:1985ze}.

To describe the medium- and long-range behavior of nuclear forces, the hybrid
quark model was developed by incorporating the exchanges of both $\pi$-meson
and $\sigma$-meson at the baryon level~\cite{Oka:1982qa}. The effective
meson exchange potential between two nucleons was introduced to simulate the
effects of the meson cloud surrounding the quark core. For consistency, the
implementation of chiral symmetry at the quark potential level was
essential~\cite{Manohar:1983md}. The constituent quark mass arises from
the spontaneous breaking of chiral symmetry at a certain momentum scale.
Once this constituent quark mass is generated, the quarks must interact
via Goldstone bosons. The $\pi$-meson and $\sigma$-meson exchanges at the
quark level were incorporated in the non-relativistic quark model, called
SU(2) ChQM. This model provides a satisfactory description of hadron spectra, nucleon-nucleon
phase shifts, and the deuteron~\cite{Obukhovsky:1990tx, Fernandez:1993hx, Valcarce:1995dm}.

Later, the SU(3) ChQM was developed to investigate interactions between nucleons
and hyperons, as well as hyperon-hyperon interactions~\cite{Fujiwara:1996qj}.
In the model, the chiral symmetry is spontaneously broken in the light sector
($u$, $d$, and $s$) while it is explicitly broken in the heavy sector ($b$ and $c$).
The interactions mediated by the $\pi$-, $K$-, $\eta$-, and $\sigma$-meson
exchange occur exclusively in the light quark sector.

It is a logical advancement to extend this model to encompass a broader range of
hadron-hadron interactions, including those involving heavy hadrons. We employed
the model to investigate single-flavored dibaryon bound states in the $^1S_0$
configuration~\cite{Deng:2023zlx}. Our findings were compared with results from
various theoretical frameworks, particularly lattice QCD. Many of our conclusions
were consistent and occasionally qualitatively aligned with those derived from
other approaches. Furthermore, the model effectively describes the characteristics
of the deuteron-like $DD^*$ molecular state $T^+_{cc}(3875)$~\cite{Deng:2021gnb}.
These results indicate that our model is robust and physically valid in its
representation of hadron-hadron interactions.

To sum up, the complete model Hamiltonian involved in this study is expressed as
\begin{widetext}
\begin{eqnarray}
\begin{aligned}
&H_n=\sum_{i=1}^n \left(m_i+\frac{\mathbf{p}_i^2}{2m_i}
\right)-T_{\rm cm}+\sum_{i<j}^n V_{ij}^{\rm oge}+V_{ij}^{\rm con}+V_{ij}^{\rm obe}+V_{ij}^{\sigma}, \\
&V_{ij}^{\rm oge}={\frac{\alpha_{s}}{4}}\boldsymbol{\lambda}_{i}
\cdot\boldsymbol{\lambda}_{j}\left({\frac{1}{r_{ij}}}-
{\frac{\boldsymbol{\sigma}_{i}\cdot\boldsymbol{\sigma}_{j}}
{6m_im_jr_0^2(\mu_{ij})r_{ij}}}e^{-\frac{r_{ij}}{r_0(\mu_{ij})}}\right),~V^{\rm con}_{ij}=-a_c\boldsymbol{\lambda}_i\cdot\boldsymbol{\lambda}_jr^2_{ij} \\
&V_{ij}^{\rm obe}=V^{\pi}_{ij}\sum_{k=1}^3\boldsymbol{\lambda}_i^k\cdot\boldsymbol{\lambda}_j^k
+V^{K}_{ij}\sum_{k=4}^7\boldsymbol{\lambda}_i^k\cdot\boldsymbol{\lambda}_j^k
+V^{\eta}_{ij}(\boldsymbol{\lambda}_i^8\cdot\boldsymbol{\lambda}_j^8\cos \theta_P
-\sin \theta_P)\\
&V^{\chi}_{ij}=\frac{g^2_{ch}}{4\pi}\frac{m^3_{\chi}}{12m_im_j}
\frac{\Lambda^{2}_{\chi}}{\Lambda^{2}_{\chi}-m_{\chi}^2}
\boldsymbol{\sigma}_{i}\cdot
\boldsymbol{\sigma}_{j}\left( Y(m_\chi r_{ij})-
\frac{\Lambda^{3}_{\chi}}{m_{\chi}^3}Y(\Lambda_{\chi} r_{ij})\right),~Y(x)=\frac{e^{-x}}{x},\\
&V^{\sigma}_{ij}=-\frac{g^2_{ch}}{4\pi}
\frac{\Lambda^{2}_{\sigma}m_{\sigma}}{\Lambda^{2}_{\sigma}-m_{\sigma}^2}
\left( Y(m_\sigma r_{ij})-
\frac{\Lambda_{\sigma}}{m_{\sigma}}Y(\Lambda_{\sigma}r_{ij})
\right).
\end{aligned}
\end{eqnarray}
\end{widetext}
$m_i$ and $\mathbf{p}_i$ represent the mass and momentum of the quark $q_i$,
respectively. $T_{\rm cm}$ denotes the center-of-mass kinetic energy. $\boldsymbol{\lambda}_{i}$
and $\boldsymbol{\sigma}_{i}$ stand for the SU(3) Gell-Mann matrices and SU(2)
Pauli matrices, respectively. $r_{ij}$ is the distance between $q_i$ and $q_j$
and $\mu_{ij}$ is their reduced mass, with the relation $r_0(\mu_{ij})=\frac{\hat{r}_0}{\mu_{ij}}$.
The quark-gluon coupling constant $\alpha_s$ adopts an effective scale-dependent
form,
\begin{equation}
\alpha_s(\mu_{ij})=\frac{\alpha_0}{\ln\frac{\mu_{ij}^2}{\Lambda_0^2}}.
\end{equation}
The symbol $\chi$ represents $\pi$, $K$ and $\eta$ meson and their mass $m_{\chi}$
take the experimentally determined values, while the cutoff parameters $\Lambda_{\chi}$
and the mixing angles $\theta_{P}$  are adopted from Ref.~\cite{Vijande:2004he}.
The mass parameter $m_{\sigma}$ can be derived from the PCAC relation
$m^2_{\sigma}\approx m^2_{\pi}+4m^2_{u,d}$~\cite{Scadron:1982eg}. The chiral coupling
constant $g_{ch}$ can be calculated from the $\pi NN$ coupling constant through the
following expression,
\begin{equation}
\frac{g_{ch}^2}{4\pi}=\left(\frac{3}{5}\right)^2\frac{g_{\pi NN}^2}
{4\pi}\frac{m_{u,d}^2}{m_N^2}.
\end{equation}

The following step is to incorporate ordinary baryons within the quark model in order
to determine the adjustable model parameters with the MINUIT program. This program
is widely used for fitting models to data, aiming to minimize an objective function and derive
the best-fit parameter values along with their associated uncertainties~\cite{James:1975dr}.
We minimize the mean square error
\begin{eqnarray}
\Delta=\sum_{i=1}^N\frac{(M_i-m_i)^2}{N}\nonumber
\end{eqnarray}
to fit the mass spectrum and to determine the adjustable parameters and their errors
in the MINUIT program. $N$ is the total number of baryons.
$M_i$ is the experimental mass of the ith baryon and $m_i$ is its predicted mass in
the model. This objective function quantifies the goodness of fit between the model
and the observed data.

The central values of the adjustable parameters, along with the heavy ground state
baryon spectrum, were detailed in Tables I and II of our previous work \cite{Deng:2023zlx}.
In the current study, we present the adjustable parameters along with their associated
uncertainties in Table~\ref{parameters}. It is noteworthy that the quark masses $m_{u,d}$
and $m_s$ are held fixed during fitting baryon spectrum. The masses $m_{u,d}$ are set to
one-third of the nucleon mass, namely $m_{u,d}=313$ MeV, while $m_s$ is fixed at 500 MeV
in this investigation.

\begin{table}[ht]
\caption{Model parameters. Quark masses and $\Lambda_0$ unit in MeV, $a_c$ unit in
MeV$\cdot$fm$^{-2}$, $r_0$ unit in MeV$\cdot$fm and $\alpha_0$ is dimensionless.}\label{parameters}
\tabcolsep=0.22cm
\begin{tabular}{lccccccccccccccccc}
\toprule[0.8pt]\toprule[0.8pt] \noalign{\smallskip}
Parameter&$m_c$&$m_b$&$a_c$  \\
\noalign{\smallskip}
\toprule[0.8pt] \noalign{\smallskip}
Value   & $1613.7\pm0.6$  & $4981.6\pm0.7$ & $45.6\pm0.5$   \\
\noalign{\smallskip}
\toprule[0.8pt] \noalign{\smallskip}
Parameter&$\alpha_0$&$\Lambda_0$&$r_0$   \\
\noalign{\smallskip}
\toprule[0.8pt] \noalign{\smallskip}
Value   &  $3.76\pm0.02$ & $21.9\pm0.2$ & $95.7\pm1.1$  \\
\noalign{\smallskip}
\toprule[0.8pt]\toprule[0.8pt]
\end{tabular}
\end{table}
\begin{table}[ht]
\caption{Mass spectra of baryon ground states unit in MeV and mass rms radius of quark
core unit in fm. PDG is the abbreviation of particle data group. The ``$\times$" denotes
that the state does not exist in the experiment. \label{baryons}}
\tabcolsep=0.18cm
\begin{spacing}{0.60}
\begin{tabular}{ccccccc}
\toprule[0.8pt]\toprule[0.8pt] \noalign{\smallskip}
\noalign{\smallskip}
Baryon&$I(J^P)$&~Mass~&~Rms~&PDG   \\
\noalign{\smallskip}
\toprule[0.8pt] \noalign{\smallskip}
$\Lambda_c$    & $0(\frac{1}{2}^+)$            &   2270.3$\pm10.6$ & 0.44  & 2286.46$\pm$0.14 \\
$\Sigma_c$     & $1(\frac{1}{2}^+)$            &   2462.9$\pm8.3$  & 0.47  & 2452.9$\pm$0.4 \\
$\Sigma^*_c$   & $1(\frac{3}{2}^+)$            &   2492.9$\pm8.3$  & 0.48  & 2518.41$\pm$0.22 \\
$\Xi_{cc}$     & $\frac{1}{2}(\frac{1}{2}^+)$  &   3635.8$\pm7.9$  & 0.43  & 3621.6$\pm$0.4 \\
$\Xi^*_{cc}$   & $\frac{1}{2}(\frac{3}{2}^+)$  &   3667.5$\pm7.5$  & 0.44  & $\times$ \\
$\Lambda_b$    & $0(\frac{1}{2}^+)$            &   5606.6$\pm10.5$ & 0.29  & 5619.57$\pm$0.16 \\
$\Sigma_b$     & $1(\frac{1}{2}^+)$            &   5814.3$\pm8.2$  & 0.31  &5810.56$\pm$0.25 \\
$\Sigma^*_b$   & $1(\frac{3}{2}^+)$            &   5825.6$\pm8.2$  & 0.32  & 5830.32$\pm$0.27 \\
$\Xi_{bb}$     & $\frac{1}{2}(\frac{1}{2}^+)$  &   10263.5$\pm7.5$ & 0.29  & $\times$ \\
$\Xi^*_{bb}$   & $\frac{1}{2}(\frac{3}{2}^+)$  &   10276.7$\pm7.3$ & 0.29  & $\times$ \\
\noalign{\smallskip}
\toprule[0.8pt]\toprule[0.8pt]
\end{tabular}
\end{spacing}
\end{table}

Based on the model parameters and their uncertainties, we can calculate the errors
of the baryon spectrum inducing by the parameter uncertainties. The model Hamiltonian can be
regarded as a function of model parameters, $H=H(x_1,...,x_6)$. The variance of the model
hamiltonian resulting from the parameter uncertainties $\delta x_i$ can be written as
\begin{eqnarray}
\delta H=\sum_{i=1}^{6}{\frac{\partial{H(x_1,...,x_6)}}{\partial{x_i}}}\delta x_i,
\end{eqnarray}
where $x_i$ and $\delta x_i$ represent the $i$-th adjustable parameter and it's uncertainty,
respectively. The energy uncertainty introduced by the parameter uncertainties can be
calculated using the formula,
\begin{eqnarray}
\delta E_B=\langle\Phi^B_{IJ}\left|\delta H\right|\Phi^B_{IJ}\rangle,
\end{eqnarray}
where $\Phi^B_{IJ}$ is the eigen vector for baryon obtained by solving Shr\"{o}dinger equation.
The energy of baryons related to this work and their  and energy uncertainty are presented in
Table~\ref{baryons}, which ranges from 7 MeV to 10 MeV.

In addition, we calculate the mass root-mean-square (rms) radius of quark core of baryons
using their eigenvectors. The mass rms radius was defined as~\cite{Silvestre-Brac:1985aip,
Silvestre-Brac:1996myf}
\begin{eqnarray}
\langle\mathbf{r}^2\rangle^{\frac{1}{2}}=\left (\sum_{i=1}^3\frac{m_i\langle(\mathbf{r}_i-
\mathbf{R}_{\rm cm})^2\rangle}{m_1+m_2+m_3}\right )^{\frac{1}{2}}.
\end{eqnarray}
The numerical results are presented in Table~\ref{baryons}, which are in good agreement with
those from~\cite{Silvestre-Brac:1985aip,Silvestre-Brac:1996myf}. While the mass rms radius
is not directly observable, it is still a valuable quantity that provides insight into the
size of baryons within the framework of the constituent quark model. Generally, the mass rms
radius of the quark core is smaller than the physical radius of the baryons, as it does not
account for the contributions of the meson cloud surrounding the valence quarks, which are
excluded from the model calculations.

\section{wave functions of dibaryons}

The wave function of the dibaryons with well-defined isospin and spin ($I$ and $J$)
can be expressed as
\begin{eqnarray}
\Psi_{IJ}^{\rm Dibaryon}=\sum_{\xi}
c_{\xi}\mathcal{A}\left\{\left[\Phi^{\rm B_1}_{I_1J_1}
\Phi^{\rm B_2}_{I_2J_2}\right]_{IJ}\phi_{lm}(\boldsymbol{\rho})\right\},
\end{eqnarray}
where $\Phi^{\rm B_1}_{I_1J_1}$ and $\Phi^{\rm B_2}_{I_2J_2}$ stand for the wave
functions of the ground state baryons $B_1$ and $B_2$, respectively. More details
about the construction of baryons can be found in our previous work~\cite{Deng:2023zlx}.
The square brackets indicate the Clebsch-Gordan couplings of spin and isospin.
The summation index $\xi$ encompasses all possible isospin-spin intermediate
configurations $\{I_1, I_2, J_1, J_2\}$ that can be coupled to yield the total
isospin $I$ and angular momentum $J$ of the dibaryons. The coefficients $c_{\xi}$
are dictated by the dynamics of the model and the inherent properties of the baryons.
They can be obtained by solving the eigen problem.

The term $\phi_{lm}(\boldsymbol{\rho})$ is the relative motion wave function between
two baryons, where
\begin{eqnarray}
\boldsymbol{\rho}=\frac{m_1\mathbf{r}_1+m_2\mathbf{r}_2+m_3\mathbf{r}_3}{m_1+m_2+m_3}
-\frac{m_4\mathbf{r}_4+m_5\mathbf{r}_5+m_6\mathbf{r}_6}{m_4+m_5+m_6}.
\end{eqnarray}
Accurate numerical calculations are crucial for fully understanding the properties
of dibaryons. The Gaussian Expansion Method (GEM) has demonstrated considerable efficacy
in addressing the few-body problem within nuclear physics~\cite{Hiyama:2003cu}. A recent
comparative study highlighted the advantages of the GEM over both the resonating group method
and the diffusion Monte Carlo method when it comes to investigating tetraquark bound states
in quark models~\cite{Meng:2023jqk}. In this study, we employ the GEM to explore
the properties of dibaryon states, thereby ensuring that our numerical results are both
accurate and reliable. Within the GEM framework, the relative motion wave function
$\phi_{lm}(\boldsymbol{\rho})$ can be expanded as a superposition of Gaussian functions
with different widths, expressed as,
\begin{eqnarray}
\phi_{lm}(\boldsymbol{\rho})&=&\sum_{n=1}^{n_{max}}c_nN_{nl}x^{l}
e^{-\nu_nx^2}Y_{lm}(\hat{\boldsymbol{\rho}}).
\end{eqnarray}
For further details on the GEM, refer to Ref.~\cite{Hiyama:2003cu}.

Note that a restriction must be imposed on the quantum numbers $\{I_1, I_2, J_1, J_2,I,J,l\}$
when the baryons $B_1$ and $B_2$ are identical particles in order to satisfy Fermi-Dirac statistics.
Their quantum numbers must satisfy the relation $J_1+J_2-J+I_1+I_2-I+l=\text{odd}$. We primarily
focus on the ground state of dibaryons, i.e., with $l=0$, as these states are more likely to form
bound states.

The present work mainly focus on the di-$\Xi_{cc}$ and di-$\Xi_{bb}$ states, uniformly denoted
as $B_1(Q_1Q_2q_3)B_2(Q_4Q_5q_6)$, where $Q_i$ stands for heavy charm and bottom quarks and $q_i$
represents up and down quarks. $\mathcal{A}$ is antisymmetrization operator acting on the identical
quarks and can be expressed as
\begin{eqnarray}
\begin{aligned}
&\mathcal{A}=\mathcal{A}_{1245}\mathcal{A}_{36},~
\mathcal{A}_{36}=\left(1-P_{36}\right),\\
\noalign{\smallskip}
&\mathcal{A}_{1245}=\left(1-P_{14}-P_{15}-P_{24}-P_{25}+P_{14}P_{25}\right),
\end{aligned}
\end{eqnarray}
where $P_{ij}$ is a permutation operator acting on the identical quarks $i$ and $j$.

\section{numerical results and discussions}

\subsection{Methodology}

Next, we turn our attention to the investigation of the dibaryons within the quark models.
By approximately exact solving the six-body Schr\"{o}dinger equation
\begin{eqnarray}
(H_6-E_6)\Psi^{\rm{Dibaryon}}_{IJ}=0
\end{eqnarray}
for the bound state problem with the Rayleigh-Ritz variational principle, we can achieve
the eigenvalues and corresponding eigenvectors of the dibaryons.

The binding energy of the dibaryons can be calculated through the formula,
\begin{equation}
	\Delta E_6=E_6\left({\rho}\right) - E_6\left({\infty}\right).
\end{equation}
$E_6\left({\rho}\right)$ is the minimum energy of the dibaryons at the optimal inter-baryon
separation $\rho$. $E_6\left({\infty}\right)$ is the sum of the energies, $E(\infty)=E_{B_1}+E_{B_2}$,
of two isolated ground state baryons in the quark model, i.e. the theoretical threshold of the dibaryons.
If $\Delta E_6\geq0$, the dibaryon is unbound and can decay into two constituent baryons $B_1$ and
$B_2$ via strong interactions. However, if $\Delta E_6<0$, the strong decay into two constituent
baryons is forbidden, and the decay can only occur via weak or electromagnetic interactions.
The binding energies of the dibaryon di-$\Xi_{cc}$ and di-$\Xi_{bb}$ with various isopsin-spin
configurations are presented in Tables~\ref{di-xicc}-\ref{di-xibb1} in the following subsections.
To provide a clearer and more intuitive representation of the binding energy, the energies of
these dibaryon states and their corresponding thresholds are displayed in Figs.~\ref{fig-xicc}
and \ref{fig-xibb}. The thresholds for the dibaryons with various isopsin-spin configurations
are indicated by dotted horizontal lines. This visualization offers a better understanding
of the relative binding energies and the positioning of each state in relation to the
respective thresholds.

The uncertainty of binding energy resulting from the parameter uncertainty can be written as
\begin{eqnarray}
\delta(\Delta E_6)=\delta E_6-\delta E_{B_1}-\delta E_{B_2}.
\end{eqnarray}
Because the same model parameters contribute to both the dibaryon and its corresponding
baryon masses, the uncertainties arising from these parameters are strongly correlated.
Consequently, when the dibaryon binding energy is calculated as their difference, these
correlated contributions partially cancel, resulting in a reduced overall uncertainty,
as illustrated in Tables~\ref{di-xicc}-\ref{di-xibb1}. Importantly, these uncertainties
reflect the statistical propagation from the model parameters, rather than an artificial
suppression of errors. As a result, this subtraction of binding energy significantly reduces
the influence of uncertainties in the model parameters and baryon spectra on the calculated
binding energies. Comprehensive error analysis indicate that the predicted dibaryon states
are quantitatively reliable in the quark model.

To clarify the binding mechanism of the bound states, we analyze the contributions from
various interactions to the binding energy $\Delta E_6$ using its eigenvector,
\begin{eqnarray}
\Delta\langle V^{\chi}\rangle&=&\langle\Psi^{\rm Dibaryon}_{IJ}|V^{\chi}|\Psi^{\rm Dibaryon}_{IJ}\rangle
-\langle\Phi^{B_1}_{I_1J_1}|V^{\chi}|\Phi^{B_1}_{I_1J_1}\rangle\nonumber\\
&-&\langle\Phi^{B_2}_{I_2J_2}|V^{\chi}|\Phi^{B_2}_{I_2J_2}\rangle,
\end{eqnarray}
where $\chi$ represents all types of interactions in the quark models. Similarly, we can also
obtain the kinetic contribution to the binding energy.
Additionally, we compute the distances between two baryons in the bound state,
\begin{eqnarray}
\begin{aligned}
\langle\boldsymbol{\rho}^2\rangle^{\frac{1}{2}}=\langle\Psi^{\rm Dibaryon}_{IJ}|\boldsymbol{\rho}^2|\Psi^{\rm Dibaryon}_{IJ}\rangle^{\frac{1}{2}}.
\end{aligned}
\end{eqnarray}
The numerical results from these calculations are presented in Tables~\ref{di-xicc} and ~\ref{di-xibb}. Comparing
with the sizes of single baryons, one can intuitively figure out the spatial configuration of the bound state.
If $\langle\boldsymbol{\rho}^2\rangle^{\frac{1}{2}}$ is obviously greater than the sum of the sizes of the two
baryons, the baryons are considered to be completely separated, forming a loose, deuteron-like configuration.
Otherwise, the dibaryon is a compact hexaquark state.

\subsection{Di-$\Xi_{cc}$ states}

\begin{table*}[ht]
\caption{The binding energy $\Delta E_6$ of the di-$\Xi_{cc}$ state and the contribution of each part in the Hamiltonian to $\Delta E_6$,
$\Delta V^{\rm con}$, $\Delta V^{\rm coul}$, $\Delta V^{\rm cm}$, $\Delta T$, $\Delta V^{\sigma}$, $\Delta V^{\pi}$, and $\Delta V^{\eta}$
are confinement term, Coulomb term, chromomagnetic term, kinetic energy, $\sigma$-, $\pi$-, and $\eta$-meson exchange term, respectively,
unit in MeV. $\langle\boldsymbol{r}^2\rangle^{\frac{1}{2}}$ represents the average size of a baryon and $\langle\boldsymbol{\rho}^2\rangle^{\frac{1}{2}}$
is the distance between two baryons, unit in fm.} \label{di-xicc}
\tabcolsep=0.26cm
\begin{spacing}{1.32}
\begin{tabular}{cccccccccccc}
\toprule[0.8pt]\toprule[0.8pt] \noalign{\smallskip}
$I(J^{P})$&Dibaryon&$\Delta E_6$&$\Delta V^{\rm con}$&$\Delta V^{\rm coul}$&$\Delta V^{\rm cm}$&$\Delta T$
&$\Delta V^{\pi}$&$\Delta V^{\eta}$&$\Delta V^{\sigma}$&$\langle\boldsymbol{r}^2\rangle^{\frac{1}{2}}$&$\langle\boldsymbol{\rho}^2\rangle^{\frac{1}{2}}$\\
\noalign{\smallskip}\toprule[0.8pt] \noalign{\smallskip}
0$(1^{+})$  & $\Xi_{cc}\Xi_{cc}$     & Unbound &&&&&&&&$0.44$&$\rightarrow\infty$\\
			& $\Xi_{cc}\Xi^*_{cc}$   & $-1.5\pm0.2$ & $-1.4$ & $-1.5$ & $-1.8$ & $10.0$ & $-2.2$ &  $0.1$ & $-4.7$ & $0.44$ & $2.37$  \\
            & $\Xi^*_{cc}\Xi^*_{cc}$ & $-3.3\pm0.2$ & $-1.1$ & $-1.5$ & $-1.4$ & $3.6$  & $3.0$  & $-0.3$ & $-5.6$ & $0.44$ & $1.87$  \\
			& Coupling               & $-7.1\pm0.1$ & $-2.1$ & $-2.1$ & $-2.2$ & $3.9$  & $3.7$  & $-0.3$ & $-8.0$ & $0.44$ & $1.40$  \\
\noalign{\smallskip}
\toprule[0.8pt]\toprule[0.8pt]
\end{tabular}
\end{spacing}
\caption{Effective potential of the state $\Xi_{cc}\Xi_{cc}$ with $0(1^+)$ when the distance
$\langle\boldsymbol{\rho}^2\rangle^{\frac{1}{2}}$ ranges from 0.50 fm to 3.00 fm, see the caption for
Table \ref{di-xicc}.} \label{unbound}
\tabcolsep=0.245cm
\begin{spacing}{1.20}
\begin{tabular}{ccccccccccccc}	
\toprule[0.8pt]\toprule[0.8pt] \noalign{\smallskip}
$I(J^{P})$&Dibaryon&$\Delta E_6$&$\Delta V^{\rm con}$&$\Delta V^{\rm coul}$&$\Delta V^{\rm cm}$&$\Delta T$
&$\Delta V^{\pi}$&$\Delta V^{\eta}$&$\Delta V^{\sigma}$&$\langle\boldsymbol{r}^2\rangle^{\frac{1}{2}}$
&$\langle\boldsymbol{\rho}^2\rangle^{\frac{1}{2}}$\\
\noalign{\smallskip}\toprule[0.8pt] \noalign{\smallskip}
\multirow{6}{*}{$0(1^{+})$} & \multirow{6}{*}{$\Xi_{cc}\Xi_{cc}$}
 &$106.8\pm0.2$ & $-17.5$  & $25.0$ & $2.9$   & $90.5$  & $-16.3$ & $1.6$  & $-13.8$  & $0.44$ &  $0.50$  \\
&&$53.3\pm0.1$  & $-1.1$   & $-6.8$ & $-2.4$  & $74.6$  & $-1.2$  & $0.1$  & $-10.0$  & $0.44$ &  $1.00$  \\
&&$27.3\pm0.1$  & $-2.4$   & $-6.7$ & $-2.0$  & $45.1$  & $0.7$   & $-0.1$ & $-7.3$   & $0.44$ &  $1.50$  \\
&&$17.7\pm0.1$  & $-2.0$   & $-5.1$ & $-1.6$  & $31.3$  & $0.8$   & $-0.1$ & $-5.7$   & $0.44$ &  $2.00$  \\	
&&$13.1\pm0.1$  & $-1.6$   & $-3.9$ & $-1.3$  & $23.8$  & $0.8$   & $-0.1$ & $-4.8$   & $0.44$ &  $2.50$  \\
&&$9.6\pm0.1$   & $-1.2$   & $-2.9$ & $-1.0$  & $17.8$  & $0.7$   & $-0.1$ & $-3.9$   & $0.44$ &  $3.00$  \\
\noalign{\smallskip}	
\toprule[0.8pt]\toprule[0.8pt]
\end{tabular}
\end{spacing}
\end{table*}

\begin{figure} [h]
\centering
\resizebox{0.48\textwidth}{!}{\includegraphics{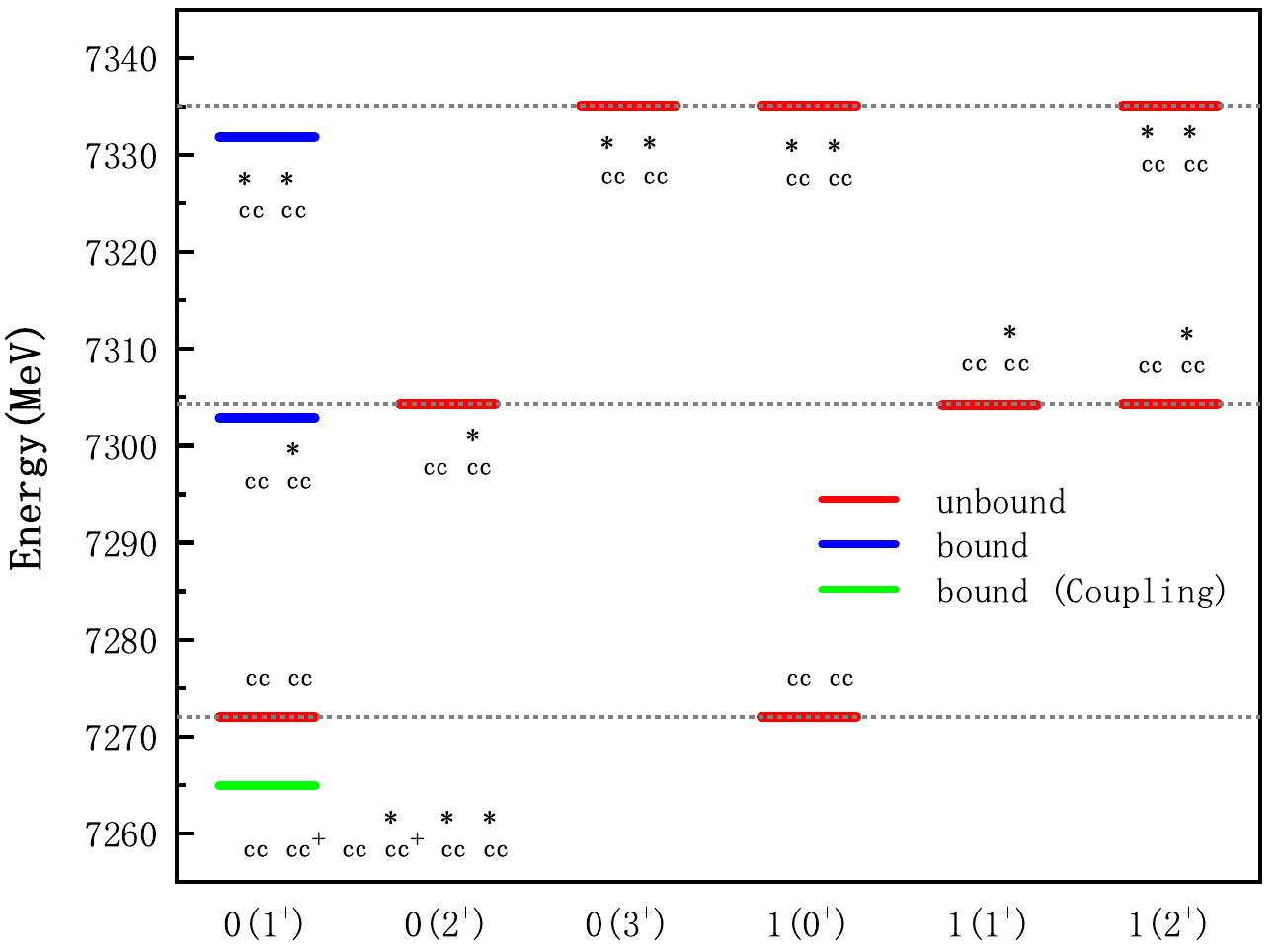}}
\caption{The energy spectrum of the di-$\Xi_{cc}$ system. The thresholds, $\Xi_{cc}\Xi_{cc}$,
$\Xi_{cc}\Xi^*_{cc}$, and $\Xi^*_{cc}\Xi^*_{cc}$, are marked by three dotted horizontal lines,
arranged from bottom to top.}
\label{fig-xicc}
\end{figure}

The dibaryon di-$\Xi_{cc}$ with $0(1^+)$ has three possible configurations: $\Xi_{cc}\Xi_{cc}$,
$\Xi_{cc}\Xi^*_{cc}$, and $\Xi^*_{cc}\Xi^*_{cc}$. From Table~\ref{di-xicc} and Fig.~\ref{fig-xicc},
it can be observed that $\Xi_{cc}\Xi_{cc}$ with $0(1^+)$ can not produce a bound state in the quark
model because of the lack of binding force. In contrast, this dibaryon configuration shows a binding
energy of approximately 0.7 MeV and a root-mean-square radius of 3.23 fm when using a reasonable
cutoff of 1.1 GeV in the OBE model~\cite{Meng:2017fwb}. This suggests that the $\Xi_{cc}\Xi_{cc}$
with $0(1^+)$ is a promising candidate for a deuteron-like molecular state in the OBE model.

The distance between two baryons in the $\Xi_{cc}\Xi_{cc}$ with $0(1^+)$ is exceptionally large,
potentially approaching infinity, which serves as a means to mitigate the repulsive interactions.
Consequently, the contributions from various interactions, as well as the kinetic energy associated
with the system, effectively diminish to nearly zero due to the large spatial separation.

The dynamical calculation involved in this work contrasts with the Born-Oppenheimer approximation employed
in molecular physics, where nuclei are assumed to be static because of their larger masses than electrons.
Similar to the Born-Oppenheimer approximation, we present the effective potential between two baryons
at specific fixed distances, as illustrated in
Table~\ref{unbound}. It is evident that the kinetic term plays a significant role in inhibiting the
formation of the bound state $\Xi_{cc}\Xi_{cc}$ with $01^+$, which is also hold truth for other unbound
dibaryon states.

Both $\Xi_{cc}\Xi^*_{cc}$ and $\Xi^*_{cc}\Xi^*_{cc}$ with $0(1^+)$ have the potential to form shallow
bound states, with binding energies of approximately 1.5 MeV and 3.3 MeV, respectively, relative to
their constituent particles in the quark model. In the OBE model, the $\Xi_{cc}\Xi_{cc}^*$ $0(1^+)$
forms a shallow bound state, with a binding energy ranging from 4.7 MeV to 12.6 MeV, depending on the
cutoff parameter $\Lambda$, which varies between 1.2 GeV and 1.6 GeV~\cite{Yang:2019rgw}. In contrast,
the $\Xi_{cc}^*\Xi_{cc}^*$ with $0(1^+)$ leads to a deep bound state, with binding energies ranging
from 25.3 MeV to 39.8 MeV when the cutoff parameter $\Lambda$ is between 0.82 GeV and 0.84 GeV~\cite{Yang:2019rgw}.

In the formation of the bound states $\Xi_{cc}\Xi^*_{cc}$ and $\Xi^*_{cc}\Xi^*_{cc}$ with $0(1^+)$
in the quark model, all the interactions $V^{\rm con}$, $V^{\rm coul}$, $V^{\rm cm}$, and $V^{\sigma}$
contribute a little attractive force, with $V^{\sigma}$ playing the dominant role, as shown in Table~\ref{di-xicc}.
In contrast, the $\pi$-meson exchange is repulsive in the $\Xi_{cc}\Xi^*_{cc}$ with $0(1^+)$, while
it is attractive in the $\Xi^*_{cc}\Xi^*_{cc}$ with $0(1^+)$. Comparatively speaking, the $\eta$-meson
exchange is very small and can almost be considered negligible. Overall contributions from those three
meson exchanges is attractive. Our numerical results indicate that once those three meson exchanges, which
just occurs between two subclusters and do not affect the corresponding threshold, are excluded from the
quark model, those two bound states disappear. This indicated that the meson exchange force plays a pivotal
role in the formation of the bound states. The main factor hindering the formation of the $\Xi^{(*)}_{cc}$
and $\Xi^*_{cc}$ subclusters into the bound dibaryon is the relative kinetic energy between two subclusters.

It is important to highlight that this processing method concerning meson exchange interactions, specifically,
evaluating its contribution by suppressing the meson exchange interaction, is applicable only to dibaryon
systems  in which each baryon consists of two heavy quarks and one light quark. Examples of such
systems include $\Xi^{(*)}_{cc}\Xi^{(*)}_{cc}$ and $\Xi^{(*)}_{bb}\Xi^{(*)}_{bb}$. However, this processing
method is not valid for other dibaryon systems that involve multiple light quarks, such as $\Lambda_c\Lambda_c$
and $\Lambda_b\Lambda_b$, as the masses of the individual hadrons would be altered when the meson exchange
interaction is turned off in the analysis of the interactions between two baryons.

The average distance $\langle\boldsymbol{\rho}^2\rangle^{\frac{1}{2}}$ between $\Xi_{cc}$ ($\Xi^*_{cc}$)
and $\Xi^*_{cc}$ is 2.37 (1.87) fm, as shown in Table~\ref{di-xicc}, which is significantly larger
than twice the size, approximately 0.44 fm of the single $\Xi^{(*)}_{cc}$ baryon, as indicated in Table~\ref{baryons}.
This suggests that two baryons are well separated and do not completely overlap, implying that both
$\Xi_{cc}\Xi^*_{cc}$ and $\Xi^*_{cc}\Xi^*_{cc}$ with $0(1^+)$ resemble a deuteron-like dibaryon in
the quark model. In the OBE model, the $\Xi_{cc}\Xi^*_{cc}$ with $0(1^+)$ behaves as a deuteron-like
dibaryon, as its root-mean-square radius lies between 1.17 fm and 2.15 fm. However, the $\Xi^*_{cc}\Xi^*_{cc}$
with $0(1^+)$ may not be an ideal candidate for a molecular state, as its root-mean-square radius is
less than 1.0 fm~\cite{Yang:2019rgw}.

From a quantum mechanical standpoint, the dibaryon states should be the linear combinations of all
possible isospin-spin configurations. The mixing of these configurations serves to push down the
energy of the states. Upon coupling the $\Xi_{cc}\Xi_{cc}$, $\Xi_{cc}\Xi^*_{cc}$, and $\Xi^*_{cc}\Xi^*_{cc}$
with $0(1^+)$, the final energy of the di-$\Xi_{cc}$ system with $0(1^+)$ is approximately $-7.1$ MeV
relative to the threshold $\Xi_{cc}\Xi_{cc}$. The coupled channel effect enhances the attraction
between two subclusters, especially for the $\sigma$ meson exchange, as shown in Table~\ref{di-xicc}.
Despite this, the di-$\Xi_{cc}$ system with $0(1^+)$ maintains a deuteronlike configuration, with the
distance between the two subclusters being approximately 1.40 fm. Its main component is $\Xi_{cc}\Xi_{cc}$,
accounting for over $99\%$, while the contributions from $\Xi_{cc}\Xi^*_{cc}$ and $\Xi^*_{cc}\Xi^*_{cc}$
are less than $1\%$. The coupled channel effect plays a vital role in the formation of the di-$\Xi_{cc}$
bound system with $0(1^+)$.

The coupled channel effect, incorporating not only S-wave but also D-wave components, in the di-$\Xi_{cc}$
system with $0(1^+)$ was considered in the OBE model~\cite{Yang:2019rgw}. This system is a deep bound state
with a binding energy ranging from 36.5 MeV to 49.3 MeV relative to the threshold $\Xi_{cc}\Xi_{cc}$, and its size
is approximately 1 fm. The S-wave components dominate the di-$\Xi_{cc}$ system, while the D-wave components
are exceedingly small.

Other isospin-spin configurations, such as $\Xi_{cc}\Xi^*_{cc}$ with $0(2^+)$, $\Xi^*_{cc}\Xi^*_{cc}$ with
$0(3^+)$, $\Xi_{cc}\Xi_{cc}$ and $\Xi^*_{cc}\Xi^*_{cc}$ with $1(0^+)$, $\Xi_{cc}\Xi^*_{cc}$ with $1(1^+)$,
and, $\Xi_{cc}\Xi^*_{cc}$ and $\Xi^*_{cc}\Xi^*_{cc}$ with $0(2^+)$ do not form any bound states in the quark
model, even when considering the coupled channel effects in the calculations, as shown in Table~\ref{di-xicc}
and Fig.~\ref{fig-xicc}. In the OBE model, the $\Xi^*_{cc}\Xi^*_{cc}$ with $1(2^+)$ is fail to form a bound
state~\cite{Yang:2019rgw}. In stark contrast, the remaining isospin-spin configurations are good candidates
of deuteronlike molecular states~\cite{Yang:2019rgw}. On the whole, the OBE model can provide more attractive
forces than the quark model in the di-$\Xi_{cc}$ system.

\subsection{Di-$\Xi_{bb}$ states}

The di-$\Xi_{bb}$ system exhibits the same isospin-spin configurations as the di-$\Xi_{cc}$ system.
In accordance with the heavy quark flavor symmetry, the properties of the di-$\Xi_{bb}$ system should be
analogous to those of the di-$\Xi_{cc}$ system. However, the significantly larger mass of $\Xi^{(*)}_{bb}$
can reduce the relative kinetic between $\Xi^{(*)}_{bb}$ and $\Xi^{(*)}_{bb}$, allowing them to
approach each other more closely. As a result, the interactions between $\Xi^{(*)}_{bb}$ and $\Xi^{(*)}_{bb}$
should become stronger. In general, this makes the di-$\Xi_{bb}$ system more prone to form a bound state
compared to the di-$\Xi_{cc}$ system.

\begin{table*}[ht]
\tabcolsep=0.26cm
\begin{spacing}{1.32}
\caption{The binding energy $\Delta E_6$ of di-$\Xi_{bb}$ state, see the caption for
Table \ref{di-xicc}.} \label{di-xibb}
\tabcolsep=0.215cm
\begin{tabular}{cccccccccccc}
\toprule[0.8pt]\toprule[0.8pt] \noalign{\smallskip}
$I(J^{P})$&Dibaryon&$\Delta E_6$&$\Delta V^{\rm con}$&$\Delta V^{\rm coul}$&$\Delta V^{\rm cm}$&$\Delta T$
&$\Delta V^{\pi}$&$\Delta V^{\eta}$&$\Delta V^{\sigma}$&$\langle\boldsymbol{r}^2\rangle^{\frac{1}{2}}$&$\langle\boldsymbol{\rho}^2\rangle^{\frac{1}{2}}$\\
\noalign{\smallskip}\toprule[0.8pt] \noalign{\smallskip}
0$(1^{+})$  & $\Xi_{bb}\Xi_{bb}$     & $-6.1\pm0.2$  & $-1.5$  & $-5.4$  & $-2.5$  &$7.8$   & $2.8$   & $-0.3$  & $-7.0$   & $0.29$  & $1.03$ \\
            & $\Xi_{bb}\Xi^*_{bb}$   & $-14.3\pm0.1$ & $-3.7$  & $-10.0$ & $-4.4$  & $9.1$  & $7.1$   & $-0.7$  & $-11.6$  & $0.29$  & $0.90$ \\
			& $\Xi^*_{bb}\Xi^*_{bb}$ & $-13.7\pm0.1$ & $-6.2$  & $-8.3$  & $-7.7$  & $28.4$ & $-7.9$  & $0.5$   & $-12.6$  & $0.29$  & $0.81$ \\
			& Coupling               & $-21.2\pm0.2$ & $-10.5$ & $-9.0$  & $-11.1$ & $80.5$ & $-60.6$ & $6.5$   & $-17.0$  & $0.29$  & $0.53$ \\
\noalign{\smallskip}
$0(2^{+})$  & $\Xi_{bb}\Xi^*_{bb}$   & $-7.4\pm0.2$  & $-1.0$  & $-5.6$  & $-1.4$  & $0.8$  & $7.2$   & $-0.5$  & $-7.0$   & $0.29$  & $1.30$ \\
\noalign{\smallskip}
$0(3^{+})$  & $\Xi_{bb}^*\Xi_{bb}^*$ & $-7.1\pm0.2$  & $-0.9$  & $-5.2$  & $-1.1$  & $0.5$  & $6.9$   & $-0.5$  & $-6.7$   & $0.29$  & $1.35$ \\
\noalign{\smallskip}
$1(0^{+})$  & $\Xi_{bb}\Xi_{bb}$     & Unbound &&&&&&&&$0.29$&$\rightarrow\infty$\\
			& $\Xi_{bb}^*\Xi_{bb}^*$ & $-0.3\pm0.1$  & $0.1$   & $-0.1$ & $1.1$   & $-1.0$ & $1.4$   & $0.3$   & $-2.2$   & $0.29$  & $3.36$ \\
        	& Coupling               & Unbound &&&&&&&&$0.29$&$\rightarrow\infty$\\
\noalign{\smallskip}
$1(1^{+})$  & $\Xi_{bb}\Xi^*_{bb}$   & Unbound &&&&&&&&$0.29$&$\rightarrow\infty$\\
\noalign{\smallskip}
$1(2^{+})$  & $\Xi_{bb}\Xi_{bb}^*$   & Unbound &&&&&&&&$0.29$&$\rightarrow\infty$\\
			& $\Xi_{bb}^*\Xi^*_{bb}$ & Unbound &&&&&&&&$0.29$&$\rightarrow\infty$\\
			& Coupling               & $-0.5\pm0.1$  &$0.1$    & $-0.1$ &  $1.6$  & $-1.7$  & $1.6$   & $0.2$   & $-2.2$   & $0.29$  & $2.79$ \\
\noalign{\smallskip}
\toprule[0.8pt]\toprule[0.8pt]
\end{tabular}

\caption{The binding energy $\Delta E_6$ of di-$\Xi_{bb}$ state in the model without meson exchanges, see the caption for
Table \ref{di-xicc}.} \label{di-xibb1}
\tabcolsep=0.235cm
\begin{tabular}{cccccccccccc}
\toprule[0.8pt] \toprule[0.8pt]\noalign{\smallskip}
$I(J^{P})$&Dibaryon&$\Delta E_6$&$\Delta V^{\rm con}$&$\Delta V^{\rm coul}$&$\Delta V^{\rm cm}$&$\Delta T$
&$\Delta V^{\pi}$&$\Delta V^{\eta}$&$\Delta V^{\sigma}$&$\langle\boldsymbol{r}^2\rangle^{\frac{1}{2}}$&$\langle\boldsymbol{\rho}^2\rangle^{\frac{1}{2}}$\\
\noalign{\smallskip}\toprule[0.8pt] \noalign{\smallskip}
0$(1^{+})$  & $\Xi_{bb}\Xi_{bb}$     & $-2.2\pm0.1$ & $-0.2$ & $-2.6$ & $-1.7$ & $2.3$  &&&& $0.29$ & $1.63$ \\
            & $\Xi_{bb}\Xi^*_{bb}$   & $-9.1\pm0.1$ & $-2.3$ & $-7.4$ & $-4.0$ & $4.7$  &&&& $0.29$ & $0.97$  \\
			& $\Xi^*_{bb}\Xi^*_{bb}$ & Unbound &&&&&&&&$0.29$&$\rightarrow\infty$  \\
			& Coupling               & $-6.7\pm0.1$ & $-1.0$ & $-7.3$ & $8.3$  & $-6.7$ &&&& $0.29$ & $0.93$  \\
\noalign{\smallskip}
$0(2^{+})$  & $\Xi_{bb}\Xi^*_{bb}$   & $-7.2\pm0.1$ & $-1.4$ & $-7.2$ & $-1.6$ & $2.9$  &&&& $0.29$  & $1.25$ \\
\noalign{\smallskip}
$0(3^{+})$  & $\Xi^*_{bb}\Xi^*_{bb}$ & $-6.8\pm0.1$ & $-1.4$ & $-6.6$ & $-1.2$ & $2.3$  &&&& $0.29$  & $1.29$ \\
\noalign{\smallskip}
$1(2^{+})$  & $\Xi_{bb}\Xi^*_{bb}$   & Unbound &&&&&&&&$0.29$&$\rightarrow\infty$  \\
			& $\Xi^*_{bb}\Xi^*_{bb}$ & Unbound &&&&&&&&$0.29$&$\rightarrow\infty$  \\
			& Coupling               & $-0.3\pm0.1$ & $0.1$  & $-0.7$ & $2.1$  & $-1.8$ &&&& $0.29$  & $3.18$ \\
\noalign{\smallskip}
\toprule[0.8pt]\toprule[0.8pt]
\end{tabular}
\end{spacing}
\end{table*}

\begin{figure} [h]
\centering
\resizebox{0.495\textwidth}{!}{\includegraphics{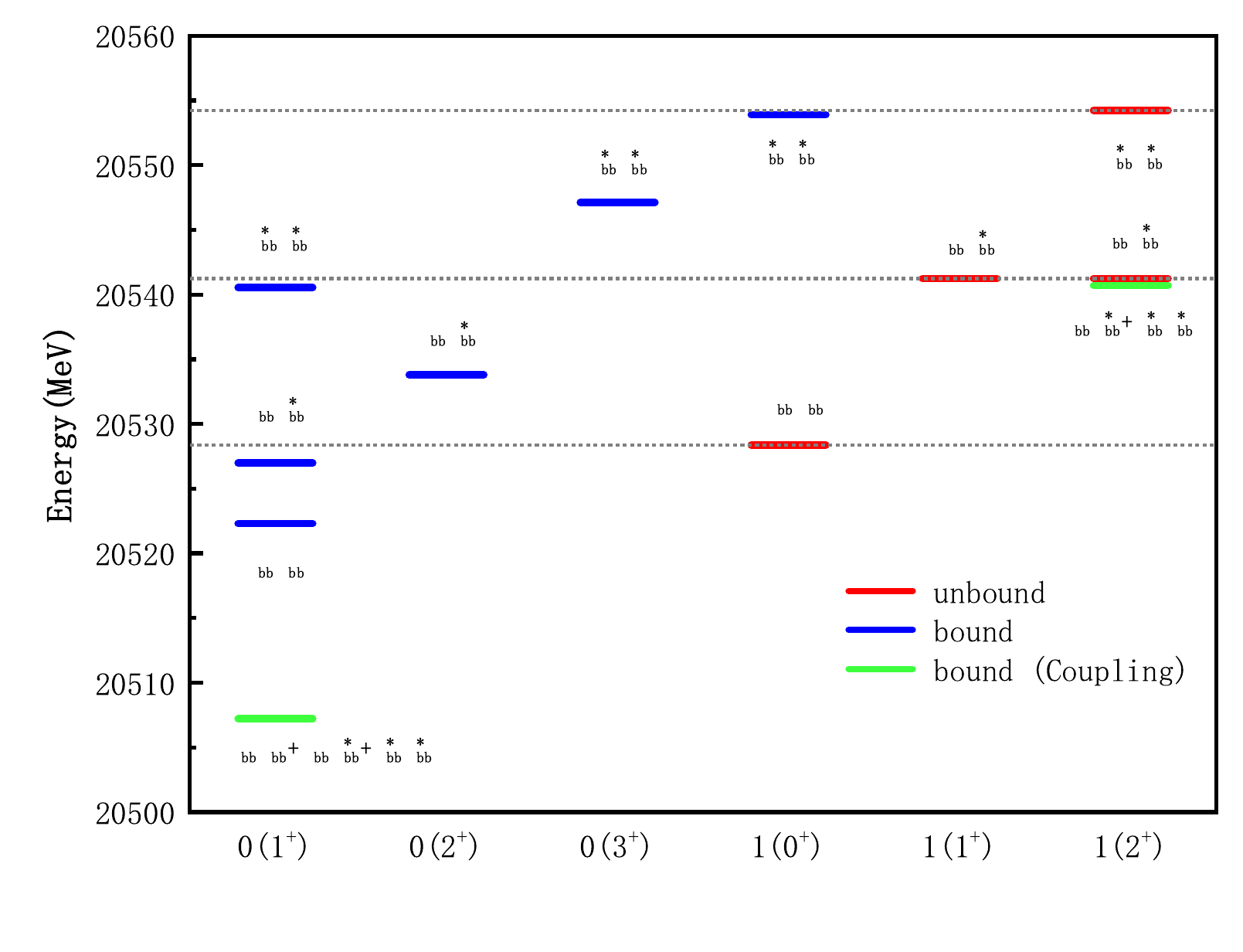}}
\caption{The energy spectrum of the di-$\Xi_{bb}$ systems. The thresholds, $\Xi_{bb}\Xi_{bb}$,
$\Xi_{bb}\Xi^*_{bb}$, and $\Xi^*_{bb}\Xi^*_{bb}$, are marked by three dotted horizontal lines,
arranged from bottom to top.}
\label{fig-xibb}
\end{figure}

From Table~\ref{di-xibb} and Fig.~\ref{fig-xibb}, it can be observed that all the $\Xi_{bb}\Xi_{bb}$, $\Xi_{bb}\Xi^*_{bb}$,
and $\Xi^*_{bb}\Xi^*_{bb}$ with $0(1^+)$ are capable of forming deuteronlike bound states within the
quark model. Their corresponding binding energies are $-6.1$ MeV, $-14.3$ MeV, and $-13.7$ MeV, respectively,
relative to the energy of their constituent particles. The distances between  $\Xi^{(*)}_{bb}$ and $\Xi^{(*)}_{bb}$
are approximately 1 fm. However, despite this small distance, their spatial configuration
does not form a compact hexaquark. Instead, they resemble a loosely bound deuteron-like molecule,
due to the small size of the $\Xi^{(*)}_{bb}$ baryons.

Similar to the di-$\Xi_{cc}$ system, all the interactions $V^{\rm con}$, $V^{\rm coul}$, $V^{\rm cm}$,
and $V^{\sigma}$ contribute a few attractive force, with $V^{\sigma}$ playing the dominant role, as
shown in Table~\ref{di-xibb}. However, the proportion of the overall contributions from those meson
exchanges to the binding energy is significantly reduced. If we turn off those meson exchanges in
the quark model, the $\Xi_{bb}\Xi_{bb}$ and $\Xi_{bb}\Xi^*_{bb}$ with $0(1^+)$ can still form deuteronlike
bound states with a binding energy of several MeVs, as shown in Table \ref{di-xibb1}. This demonstrates
that the meson exchange forces are secondary in the formation of those two bound states. However, the
$\Xi^*_{bb}\Xi^*_{bb}$ with $0(1^+)$ cannot yield a bound state when turning off those meson exchanges
in the quark model. This suggests that the meson exchanges play a critical role in the formation of
bound state $\Xi^*_{bb}\Xi^*_{bb}$ with $0(1^+)$.

After coupling of the $\Xi_{bb}\Xi_{bb}$, $\Xi_{bb}\Xi^*_{bb}$, and $\Xi^*_{bb}\Xi^*_{bb}$ with $0(1^+)$,
the final binding energy is $-21.2$ MeV relative to the threshold $\Xi_{bb}\Xi_{bb}$ in the quark model,
as illustrated in Table~\ref{di-xibb} and Fig.~\ref{fig-xibb}. This bound state is a compact hexaquark
state because two baryons are partly overlapped, which is evidenced by the sizes of baryons $\Xi^{(*)}_{bb}$
and the distance between two baryons. The probabilities of the $\Xi_{bb}\Xi_{bb}$, $\Xi_{bb}\Xi^*_{bb}$,
and $\Xi^*_{bb}\Xi^*_{bb}$ components are $41\%$, $42\%$ and $17\%$, respectively, in this compact
hexaquark state. The binding force of this compact state primarily arises from the meson exchanges,
particularly the $\pi$-meson exchange, as shown in Table~\ref{di-xibb}. Therefore, the di-$\Xi_{bb}$
with $0(1^+)$ can only form a shallow bound state with a binding energy of approximately $-6.7$ MeV,
when even considering the coupled channel effect $\Xi_{bb}\Xi_{bb}$, $\Xi_{bb}\Xi^*_{bb}$, and
$\Xi^*_{bb}\Xi^*_{bb}$ in the quark model, excluding the meson exchanges, as shown in Table~\ref{di-xibb1}.

Both $\Xi_{bb}\Xi^*_{bb}$ with $0(2^+)$ and $\Xi^*_{bb}\Xi^*_{bb}$ with $0(3^+)$ can form a deuteronlike
bound state with a binding energy of around $-7.0$ MeV and a size of approximately 1.30 fm, whether or not
the meson exchanges are involved in the quark model, as illustrated in Tables~\ref{di-xibb} and \ref{di-xibb1}.
This result arises because the contributions from the $\pi$ and $\sigma$ meson exchanges nearly cancel
each other out, while the contribution from the $\eta$ meson exchange is negligible.

The $\Xi_{bb}\Xi_{bb}$ with $1(0^+)$ is unbound in the quark model, as shown in Table~\ref{di-xibb} and
Fig.~\ref{fig-xibb}. However, the $\Xi^*_{bb}\Xi^*_{bb}$ with $1(0^+)$ can form a deuteronlike bound state,
with a binding energy of $-0.3$ MeV relative to the threshold $\Xi^*_{bb}\Xi^*_{bb}$. Its dominant binding
force arises from the $\sigma$ meson exchange. If the meson exchanges are excluded from the quark model,
this bound state ceases to exist. Additionally, the kinetic energy also contributes a slight attraction,
approximately $-1.0$ MeV. The reduction in the kinetic energy arises from the negative exchange kinetic
term~\cite{Nzar:1990ci}, which is referred to as the hadron covalent bond~\cite{Deng:2023zlx}. The coupled
channel effect between $\Xi_{bb}\Xi_{bb}$ and $\Xi^*_{bb}\Xi^*_{bb}$ does not push down the energy of the
$\Xi_{bb}\Xi_{bb}$ with $1(0^+)$ below its threshold.

The $\Xi_{bb}\Xi^*_{bb}$ with $1(1^+)$, $\Xi_{bb}\Xi^*_{bb}$ and $\Xi^*_{bb}\Xi^*_{bb}$ with $1(2^+)$ can not form
any bound states relative to their corresponding threshold in the quark model. After coupling $\Xi_{bb}\Xi^*_{bb}$
and $\Xi^*_{bb}\Xi^*_{bb}$ with $1(2^+)$, this system can establish a deuteronlike bound state with a binding
energy of approximately $-0.5$ MeV and a large size of 2.79 fm. The dominant component of this bound state
is $\Xi_{bb}\Xi^*_{bb}$, contributing $98\%$, while the $\Xi^*_{bb}\Xi^*_{bb}$ component is minimal, making
up only $2\%$. In this bound state, the meson exchanges only contribute a little attraction of $-0.4$ MeV, as
shown in Table~\ref{di-xibb}. This bound state  persists even when switching off the meson exchanges. This
suggests that the primary mechanism for the binding is not the meson exchanges, but rather the reduction in
the kinetic energy, i.e. the so-called hadron covalent bond~\cite{Deng:2023zlx}.

\section{summary}

Inspired by the $T^+_{cc}(3875)$ signal discovered by the LHCb Collaboration, we systematically examine
the properties of the heavy dibaryons di-$\Xi_{cc}$ and di-$\Xi_{bb}$ across various isospin-spin
configurations in the nonrelativistic quark model. In the calculation, we employ the Gaussian
expansion method, a high-precision numerical method, to solve the six-body Schr\"{o}dinger equations
exactly. We analyze their binding energies and spatial structures. By decomposing the attractions
from various sources, we explore the dynamical effects that govern the formation of stable bound
states under strong interactions.

The $\Xi_{cc}\Xi_{cc}$ with $0(1^+)$ can not form a bound state in the quark model. Both $\Xi_{cc}\Xi^*_{cc}$
and $\Xi^*_{cc}\Xi^*_{cc}$ with $0(1^+)$ can form deuteronlike bound states, with binding energies of
approximately $-1.5$ MeV and $-3.3$ MeV and sizes of 2.37 fm and 1.87 fm, respectively. The $\sigma$
meson exchange plays a decisive role in their formation. The coupled channel effect of those three configurations
with $0(1^+)$ enhances the attractions between two subclusters, especially for the $\sigma$ meson exchange.
The di-$\Xi_{cc}$ system with $0(1^+)$ maintains a deuteronlike configuration, with the binding energy of $-7.5$
MeV and the size of approximately 1.40 fm. Its predominant component is $\Xi_{cc}\Xi_{cc}$, accounting for
over $99\%$, while the contributions from $\Xi_{cc}\Xi^*_{cc}$ and $\Xi^*_{cc}\Xi^*_{cc}$ are less than $1\%$.
Other isospin-spin configurations, such as $0(2^+)$, $0(3^+)$, $1(0^+)$, $1(1^+)$ and $1(2^+)$, of the di-$\Xi_{cc}$
system can not establish any bound states in the quark model.

The $\Xi_{bb}\Xi_{bb}$, $\Xi_{bb}\Xi^*_{bb}$ and $\Xi^*_{bb}\Xi^*_{bb}$ with $0(1^+)$ can establish
a deuteronlike bound state, with binding energies of $-6.1$ MeV, $-14.3$ MeV and $-13.7$ MeV and sizes
of 1.03 fm, 0.90 fm and 0.81 fm in the quark model, respectively. When these three configurations are
coupled, the di-$\Xi_{bb}$ system with $0(1^+)$ is a compact hexaquark state, with a binding energy of
$-21.2$ MeV and a size of 0.53 fm. The probabilities of the $\Xi_{bb}\Xi_{bb}$, $\Xi_{bb}\Xi^*_{bb}$,
and $\Xi^*_{bb}\Xi^*_{bb}$ components are approximately $41\%$, $42\%$ and $17\%$, respectively, in
this compact hexaquark state. The binding force of this compact state primarily arises from the meson
exchanges, particularly the $\pi$-meson exchange.

Both $\Xi_{bb}\Xi^*_{bb}$ with $0(2^+)$ and $\Xi^*_{bb}\Xi^*_{bb}$ with $0(3^+)$ can form a deuteronlike
bound state with a binding energy of around $-7.0$ MeV and a size of approximately 1.30 fm, whether or not
the meson exchanges are involved in the quark model. The $\Xi_{bb}\Xi_{bb}$ with $1(0^+)$ is unbound while
the $\Xi^*_{bb}\Xi^*_{bb}$ with $1(0^+)$ can form a deuteronlike bound state, with a binding energy of $-0.3$
MeV and a size of 3.36 fm. Its dominant binding force arises from the $\sigma$ meson exchange. Additionally,
the hadron covalent bond also contributes a slight attraction, approximately $-1.0$ MeV.

The $\Xi_{bb}\Xi^*_{bb}$ with $1(1^+)$, $\Xi_{bb}\Xi^*_{bb}$ and $\Xi^*_{bb}\Xi^*_{bb}$ with $1(2^+)$
can not form any bound states in the quark model. After coupling $\Xi_{bb}\Xi^*_{bb}$ and $\Xi^*_{bb}\Xi^*_{bb}$
with $1(2^+)$, this system can establish a deuteronlike bound state with a binding energy of approximately
$-0.5$ MeV and a large size of 2.79 fm. The dominant component of this bound state is $\Xi_{bb}\Xi^*_{bb}$,
contributing $98\%$, while the $\Xi^*_{bb}\Xi^*_{bb}$ component is minimal, making up only $2\%$. Its primary
binding mechanism is not the meson exchanges, but rather the hadron covalent bond.

The information on the heavy dibaryon states presented in this work is expected to enhance our understanding
of their properties and structures from a phenomenological perspective. We also advocate various theoretical
frames to comprehensively explore the properties of those dibaryon states. Those works are attempted to
contribute meaningfully to the the hadron physics by bridging the gap between theoretical predictions and
experimental realizations. As experimental facilities such as the LHCb and forthcoming high-luminosity colliders
push the boundaries of hadron physics, the discovery of those dibaryon states may soon be attainable, despite the
challenges associated with their production in experiments.

\acknowledgments{}

This research is supported by the National Natural Science Foundation of China under Grants No. 12305096,
Chongqing Natural Science Foundation under Project No. CSTB2025NSCQ-GPX0516, and the Fundamental Research
Funds for the Central Universities under Grant No. SWU-XDJH202304 and No. SWU-KQ25016.

\end{document}